# The Nazca Frigatebird and Fomalhaut

**Amelia Carolina Sparavigna**
Department of Applied Science and Technology
Politecnico di Torino, C.so Duca degli Abruzzi 24, Torino, Italy

It seems that the Nazca Lines, ancient geoglyphs of a Peruvian dry pampa, had some astronomical orientations. Here we discuss a geoglyph representing a frigatebird and its orientation towards the setting of Fomalhaut, using freely available software.

Key-words: Geoglyphs, Nazca Lines, Orientation, Archaeoastronomy, Stellarium

The most famous geoglyphs of Peru are the Nazca Lines. Considered as one of the mysteries of the ancient world, the lines have been included among the UNESCO World Heritage Sites. Located in a large region between the towns of Nazca and Palpa, these geoglyphs are huge stylized drawings of animals ranging in size up to 300 m. Besides these drawings, there are also trapezoidal lines, spirals and dots [1]. The Nazca dry region is therefore hosting a unique remain of one of the civilizations that flourished in Peru before the arrival of Spanish people.
It is generally believed that the Lines were created by the so-called Nazca culture between 400 and 650 CE. In [2], authors reported the use of optically stimulated luminescence for dating of quartz buried when the lines were constructed. They suggest that the lines were made in the later part of the Early Intermediate Period by people of the Nazca Culture, which flourished from 100 to 800 CE.
The geoglyphs were made on the dry pampa by removing the uppermost surface, exposing the underlying ground which has a different color. This technique produces a "negative" geoglyph. A "positive" geoglyphs is instead created by the arrangement of stones, gravel or earth [3]. According to Ref.4, there are several interpretations on the purpose of these geoglyphs, but a religious intent is generally attributed to them. The simple geometric lines could be connected with the flowing of water. Some animals could be symbols of fertility. But, as told in [4], some studies are proposing that the Lines were giant astronomical calendars [5].
Let us assume that the Nazca Lines could have some links with astronomical orientations. A complete analysis of the lines is beyond the aim of this paper: here we just analyze one of the geoglyphs, to show how we can apply freely available software (the Stellarium, a software planetarium) in determining orientations. In Fig.1 we see the geoglyph that we will discuss: it is representing a Frigatebird having a very long bill. Because of its huge size, it is better recognized from the satellite imagery.
The frigatebirds are large birds with iridescent black feathers. They have long wings, tails and beaks and the males have a red pouch that is inflated to attract females during the mating season. Frigatebirds are found over tropical oceans: they cannot swim and are not able to walk well or take off from a flat surface. They are then essentially seabirds, landing only on cliffs.
Feeding habits of frigatebirds are pelagic, obtaining food by snatching it from the ocean surface, using their long, hooked bills. The geoglyph that we see in Fig.1 has such a bill and the pouch. Of course, ancient people of Nazca knew very well the habits of these birds, observing them flying over the ocean. Therefore, it is natural to ask ourselves about the bird represented by the geoglyph of Nazca, what is the Frigatebird snatching from the horizon? In my opinion, it could be Fomalhaut, the eye of the Piscis Austrinus.

Piscis Austrinus or Australis was one of the constellations listed by Ptolemy, and it also one of the modern constellations. Besides by the Greek mythology, this constellation was present also in the Egyptian astronomy. Fomalhaut (Alpha Piscis Austrini) is one of the brightest stars in the sky and its name derives from Arabic fum al-ḥawt, meaning "mouth of the fish".

First of all, let us see what the astronomic direction of the Nazca Frigatebird is. We can measure the direction the hooked bill forms with respect the cardinal West-East direction and compare with setting of celestial bodies. The measured angle of the Frigatebird from satellite images is 35°. This angle is larger than the highest azimuth of sunsets on solstices [6].

Assuming the West-East axis to measure the azimuth Z, we can define it as $Z = 90° - \arccos(\sin D/\cos L)$, where D is the declination having the highest value of 23.5° and L the latitude. The highest value of Z for the Sun at Nazca is of about 24°. For a Moon standstill, it is of 30°. Therefore the Frigatebird is not aligned with Sun or Moon settings.

To see what could be the celestial body setting with an azimuth Z angle of 35°, we can use several approaches. Here we propose the freely available planetarium software, Stellarium. This software shows a realistic sky in 3D, "just like what you see with the naked eye, binoculars or a telescope." [7] It is possible to insert coordinates and time, use an azimuthal or equatorial grid, see constellations and the names of the main celestial bodies. In this manner, using Stellarium, we see that at Nazca, approximately 30° from West it is setting Fomalhaut (see the snapshots in Fig.2). Instead of using the present time, we can set a date of 1000 CE: Fomalhaut was setting approximately at 35° degrees, corresponding to the direction of the Frigatebird geoglyph. To have a more precise result, measurements of the setting of the star directly on the site is necessary. In any case, Stellarium is free and easy for the simulations of possible alignments of ancient sites.

As we have previously told, Fomalhaut is a bright star that had a role in the lore of ancient people. It is therefore probable that it was the same for the ancient Peruvians. Since they represented a pelagic bird, it could be that Fomalhaut was the eye of an imaginary fish for the ancient Nazca people too.

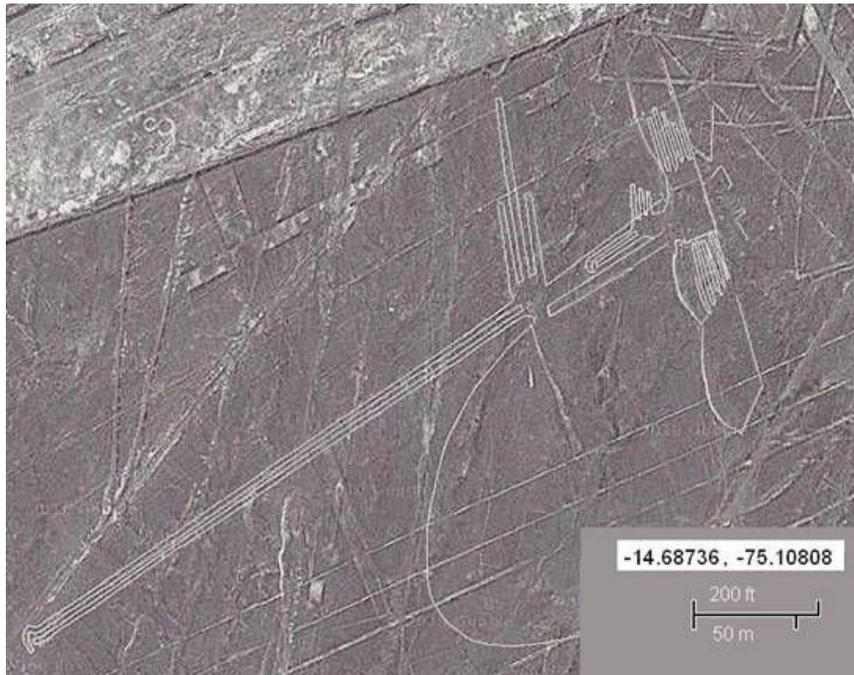

Fig.1 The geoglyphs of a frigatebird and a heron at Nazca. Note the long beak. Is the frigatebird snatching anything from the horizon? It could be Fomalhaut, the eye of a Fish.

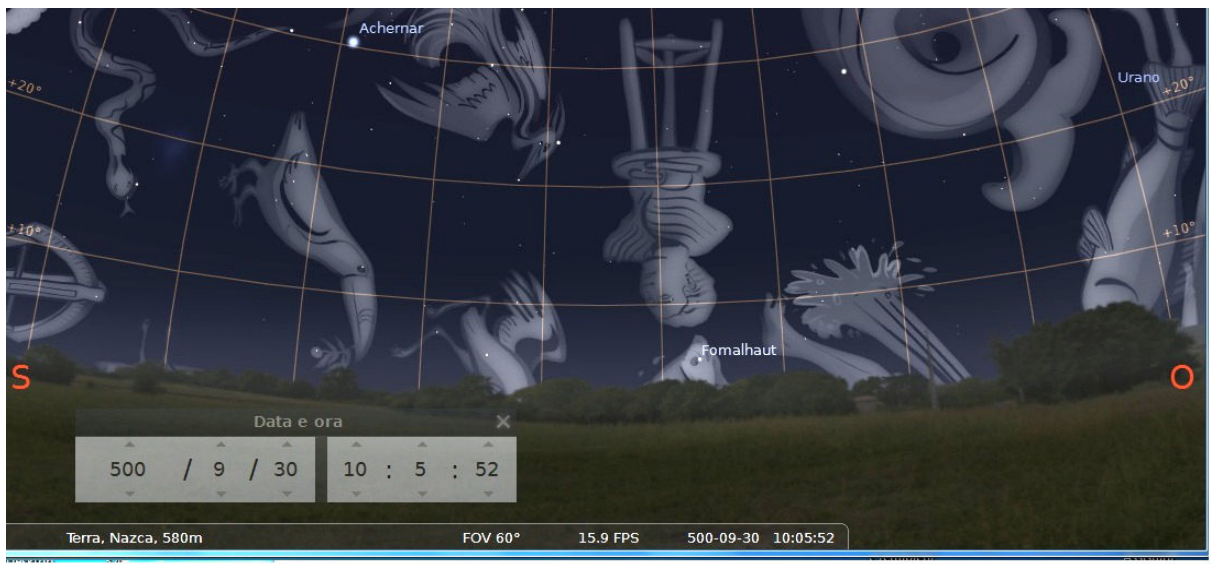

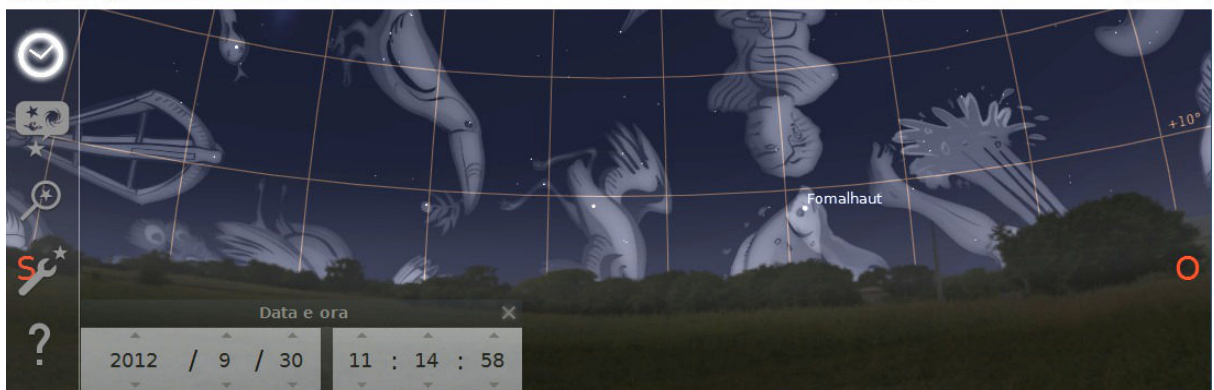

Fig.2 Two snapshots of the setting of Fomalhaut at the coordinates of Nazca on Stellarium. O is Ovest, that is West. The upper part shows the setting on 500 CE, the lower part the setting on 2012. The precession of Earth changes the azimuth.